\documentclass[%
 reprint,
 amsmath,amssymb
]{revtex4-2}

\usepackage{multirow}
\usepackage{graphicx}
\usepackage{dcolumn}
\usepackage[toc,page]{appendix}
\usepackage{bm}
\usepackage{amsmath}
\usepackage[english]{babel}
\usepackage{xcolor}
\usepackage{hyperref}
\usepackage{times}
\usepackage{tikz}
\usepackage[normalem]{ulem}
\usepackage{tikz-3dplot}
\hypersetup{
    citecolor=blue,
    colorlinks=true,
    linkcolor=blue,
    filecolor=blue,      
    urlcolor=blue,
}

\usepackage{chngcntr}
\counterwithin{table}{section}

\usepackage[english]{babel} 

\begin{document}

\title{Spin-orbit torque control of spin waves in a ferromagnetic waveguide}

\author{Andrei I. Nikitchenko}
\author{Nikolay A. Pertsev$^*$}
\affiliation{Ioffe Institute, 194021 St. Petersburg, Russia\\*pertsev.domain@mail.ioffe.ru}

\begin{abstract}
    Spin-orbit torque (SOT) created by a spin current injected into a ferromagnet by an adjacent heavy metal or topological insulator represents an efficient tool for the excitation and manipulation of spin waves. Here we report the micromagnetic simulations describing the influence of SOT on the propagation of spin waves in the $\mathrm{W}/\mathrm{CoFeB}/\mathrm{MgO}$ nanostructure having voltage-controlled magnetic anisotropy (VCMA). The simulations show that two spin waves travelling in the opposite directions can be generated in the center of the $\mathrm{CoFeB}$ waveguide via the modulation of VCMA induced by a microwave voltage locally applied to the $\mathrm{MgO}$ nanolayer. The amplitudes of these waves exponentially decrease with the propagation distance with similar decay lengths of about 2.5~$\mu$m. In the presence of a direct electric current injected into the $\mathrm{W}$ film beneath the waveguide center, the decay lengths of two spin waves change in the opposite way owing to different directions of the electric currents flowing in the underlying halves of the $\mathrm{W}$ layer. Remarkably, above a critical current density (about $2 \times 10^{10}$~A~m$^{-2}$ at zero absolute temperature and $19 \times 10^{10}$~A~m$^{-2}$ at room temperature), SOT provides the amplification of the spin wave propagating in one half of the waveguide and strongly accelerates the attenuation of the wave travelling in the other half. As a result, a long-distance spin-wave propagation takes place in a half of the $\mathrm{CoFeB}$ waveguide only. Furthermore, by reversing the polarity of the dc voltage applied to the heavy-metal layer one can change the propagation region and switch the travel direction of the spin wave in the ferromagnetic waveguide. Thus, the $\mathrm{W}/\mathrm{CoFeB}/\mathrm{MgO}$ nanostructure can be employed as an electrically controlled magnonic device converting the electrical input signal into a spin signal, which can be transmitted to one of two outputs of the device.
\end{abstract}

\maketitle

\setlength{\parindent}{15pt}
\setlength{\parskip}{0pt}

\section{Introduction \label{sec:intro}}
Generation and propagation of spin waves in magnetic nanostructures currently attract great attention because such waves can be employed for the development of energy-efficient nanodevices for information transmission and processing~\cite{Grundler2009, Chumak2015, Csaba2017}. The traditional technique of spin-wave excitation uses a microwave magnetic field created by a microstrip antenna~\cite{Demidov2011, Yu2016, Liu2018}, but it suffers from relatively low conversion efficiency. During the last decade, novel excitation techniques have been proposed and developed, which utilize spin-polarized electric currents exerting spin-transfer torque on the magnetization~\cite{Demidov2010, Madami2011, Zhou2019}, voltage-controlled magnetic anisotropy (VCMA) in ferromagnet-dielectric heterostructures~\cite{Verba2014}, and spin-orbit torque (SOT) generated by a spin current injected into a ferromagnetic film by an adjacent heavy metal~\cite{Divinskiy2018, Fulara2019}. The approaches based on SOT are especially promising, because they enable not only the excitation of spin waves but also a strong increase in their propagation length~\cite{An2014, Evelt2016, Wimmer2019, Navabi2019, Demidov2020}. Such an increase results from partial compensation of magnetic damping by SOT, which is created at the ferromagnet boundary by an electric current flowing in the heavy metal or topological insulator with a strong spin-orbit interaction via the spin Hall effect~\cite{Manchon2019}.

The functioning of magnonic devices also requires efficient control and manipulation of propagating spin waves~\cite{Mahmoud2020}. In a spin-wave (magnon) transistor, the flow of spin waves from source to drain is modulated by spin waves injected from the gate~\cite{Chumak2014, Balinskiy2018}. A spin-wave multiplexer or demultiplexer operates by guiding spin waves into one arm of Y- or T-shaped structures with the aid of magnetic fields~\cite{Vogt2014, Sadovnikov2015, Davies2015}. The dipolar interaction between the two laterally adjacent spin-wave waveguides makes it possible to develop a reconfigurable nanoscale spin-wave directional coupler, which can function as a multiplexer, tunable power splitter, or frequency separator~\cite{Wang2018}. Since SOT can strongly affect the damping of spin waves~\cite{Hamadeh2014}, it could be useful for their control and modulation as well.

\begin{figure}
    \center{\includegraphics[width=1.03\linewidth]{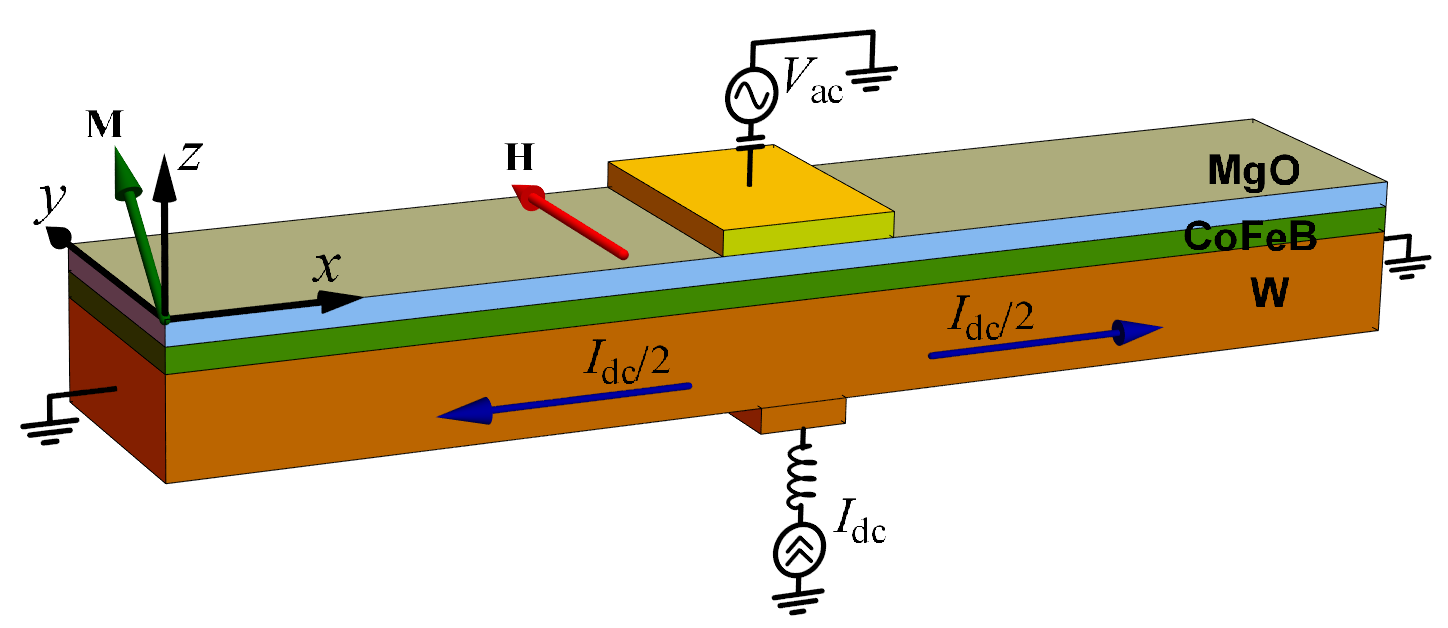}}
    \caption{\label{fig:structure} Schematic representation of $\mathrm{W}/\mathrm{CoFeB}/\mathrm{MgO}$ heterostructure subjected to a microwave voltage $V_\mathrm{ac}$ locally applied to the $\mathrm{MgO}$ nanolayer. The magnetization $\mathbf{M}$ is inclined with respect to the $\mathrm{CoFeB}$ surfaces owing to the perpendicular anisotropy associated with the $\mathrm{CoFeB}|\mathrm{MgO}$ interface and the in-plane magnetic field $\mathbf{H}$. The microwave voltage excites two spin waves in the $\mathrm{CoFeB}$ waveguide, which propagate in the opposite directions from the excitation area. The direct electric current $I_\mathrm{dc}$ injected into the $\mathrm{W}$ film changes the propagation lengths of these waves in the opposite way due to different directions of the charge flow in two halves of the $\mathrm{W}$ layer.}
\end{figure}

In this work, we propose a spin-wave switch controlled by SOT created by a direct electric current flowing in a heavy-metal layer adjacent to a ferromagnetic waveguide. In such a device, two spin waves are excited at the center of the waveguide, which propagate in opposite directions and have similar decay lengths in the absence of SOT. In contrast, when sufficient non-uniform SOT is created by an electric current injected into the heavy-metal layer near the waveguide center, the propagation length of one of these waves strongly increases, whereas the other wave experiences a fast decay. We validate our proposal by micromagnetic simulations performed for the $\mathrm{W}$/$\mathrm{CoFeB}$/$\mathrm{MgO}$ heterostructure, where spin waves are generated electrically by an oscillating VCMA associated with the $\mathrm{CoFeB}|\mathrm{MgO}$ interface (Fig.~\ref{fig:structure}). In particular, it is shown that, at zero absolute temperature, the critical current density providing complete compensation of magnetic damping for one of the generated spin waves amounts to $2 \times 10^{10}$~A~m$^{-2}$ only. At this density, the current-induced SOT reduces the decay length of the spin wave propagating in the other half of the waveguide from 2.5 to 1.2~$\mu$m. Moreover, the temporary spin-wave amplification by SOT could take place even at room temperature, but it requires the current density exceeding $19 \times 10^{10}$~A~m$^{-2}$.

\section{MODELING OF ELECTRICAL EXCITATION \\ AND CONTROL OF SPIN WAVES}
To determine the dynamics of the magnetization $\mathbf{M}(\mathbf{r}, t)$ in the ferromagnetic layer modeled by an ensemble of nanoscale computational cells, we numerically solve the modified Landau-Lifshitz-Gilbert equation, which in the presence of SOT takes the form~\cite{Sklenar2016}
 
\begin{equation}
    \begin{gathered}
    \displaystyle\frac{d\mathbf{m}}{dt} = -\gamma \mu_0 \mathbf{m} \times \mathbf{H}_\mathrm{eff} + \alpha \mathbf{m} \times \displaystyle\frac{d\mathbf{m}}{dt} \\
    + \tau_\mathrm{FL} \mathbf{s} \times \mathbf{m} + \tau_\mathrm{DL} \mathbf{m} \times (\mathbf{s} \times \mathbf{m}),
    \end{gathered}
    \label{eq:LLG}
\end{equation}
\noindent
where $\mathbf{m} = \mathbf{M} / M_s$ is the unit vector parallel to the magnetization of the cell, $M_s$ is the saturation magnetization regarded as a constant quantity at a given temperature, $\mathbf{H}_\mathrm{eff}$ is the effective field acting on the magnetization, $\mathbf{s}$ is the unit vector parallel to the spins flowing from the heavy-metal layer into the ferromagnetic film, $\gamma$ is the gyromagnetic ratio, $\mu_0$ is the magnetic permeability of free space, $\alpha = \alpha_0 + \delta \alpha$~ is the effective Gilbert damping parameter \cite{Tserkovnyak2002}, and $\tau_\mathrm{FL}$ and $\tau_\mathrm{DL}$ stand for the coefficients of field-like and damping-like SOTs. The effective field $\mathbf{H}_\mathrm{eff}$ in our case is defined by the relation $\mathbf{H}_\mathrm{eff} = \mathbf{H} + \mathbf{H}_\mathrm{ex} + \mathbf{H}_\mathrm{dip} + \mathbf{H}_\mathrm{an} + \mathbf{H}_\mathrm{PMA} + \mathbf{H}_\mathrm{DMI}$, where $\mathbf{H}$ is the external magnetic field, $\mathbf{H}_\mathrm{ex}$ and $\mathbf{H}_\mathrm{dip}$ are the contributions resulting from the exchange and dipolar interactions between spins in $\mathrm{CoFeB}$, $\mathbf{H}_\mathrm{an}$ accounts for the influence of the bulk-like cubic anisotropy of $\mathrm{CoFeB}$, while $\mathbf{H}_\mathrm{PMA}$ and $\mathbf{H}_\mathrm{DMI}$ are the effective fields caused by the perpendicular magnetic anisotropy (PMA) associated with the $\mathrm{CoFeB}|\mathrm{MgO}$ interface and the interfacial Dzyaloshinskii-Moriya interaction (DMI) at the $\mathrm{W}|\mathrm{CoFeB}$ contact, respectively. Since we consider ultrathin $\mathrm{CoFeB}$ films that can be modeled using only one computational cell in the thickness direction $z$, the interfacial contributions $\mathbf{H}_\mathrm{PMA}$ and $\mathbf{H}_\mathrm{DMI}$ are taken to be inversely proportional to the $\mathrm{CoFeB}$ thickness $t_\mathrm{F}$. The field $\mathbf{H}_\mathrm{ex}$ was evaluated via the summation of the exchange interactions of the considered inner cell with its four nearest neighbors in the film plane using the exchange constant $A_\mathrm{ex}$ introduced in the continuum approximation~\cite{Kittel1949}. The dipolar field $\mathbf{H}_\mathrm{dip}$ acting on each cell was found by summing the magnetic fields created by all other cells modeled by uniformly magnetized rectangular prisms~\cite{Azovtsev2016}. The anisotropy field $\mathbf{H}_\mathrm{an} = -(\mu_0 M_s)^{-1} \partial F_\mathrm{an} / \partial \mathbf{m}$ was determined via the differentiation of the fourth-order terms $K_1 (m_x^2m_y^2+m_x^2m_z^2+m_y^2m_z^2)$ in the expansion of the energy $F_\mathrm{an}(\mathbf{m})$ of magnetocrystalline anisotropy  because sixth-order terms are negligible in the case of $\mathrm{CoFeB}$~\cite{Ikeda2010}. As the PMA caused by the $\mathrm{CoFeB}|\mathrm{MgO}$ interface linearly depends on the electric field $E_z$ in the MgO layer~\cite{Alzate2014}, the field $\mathbf{H}_\mathrm{PMA}$ can be written as $H_z^\mathrm{PMA}= - 2 (\mu_0 M_s t_\mathrm{F})^{-1} (K_s^0+k_s E_z)m_z$, where $K_s^0=K_s (E_z=0)$, and $k_s=\partial K_s/ \partial E_z$ is the electric-field sensitivity of $K_s$~\cite{PertsevSciReps2013}. Finally, the DMI contribution to $\mathbf{H}_\mathrm{eff}$ was evaluated using the discretized version of the relation~\cite{David2018, Perez2014, Rohart2013, Vandermeulen2016}
\begin{equation}
    \mathbf{H}_\mathrm{DMI}=-\frac{D}{\mu_0 M_s t_\mathrm{F}} \bigg[ \frac{\partial m_z}{\partial x} \mathbf{e}_x + \frac{\partial m_z}{\partial y} \mathbf{e}_y - \bigg( \frac{\partial m_x}{\partial x}  +  \frac{\partial m_y}{\partial y}  \bigg) \mathbf{e}_z \bigg],
\end{equation}
\noindent
where $D$ denotes the strength of interfacial DMI, and $\mathbf{e}_i$ $(i = x, y, z)$ are the unit vectors parallel to the coordinate axes $x$, $y$, and $z$. For the computational cells adjacent to the waveguide lateral free surface $\Gamma$, the fields $\mathbf{H}_\mathrm{ex}$ and $\mathbf{H}_\mathrm{DMI}$ were calculated using the boundary conditions~\cite{Rohart2013}
\begin{equation}
    \begin{gathered}
        \frac{\partial m_z}{\partial x}\bigg|_\Gamma = \frac{D}{2 A_\mathrm{ex}} m_x,\\
        \frac{\partial m_z}{\partial y}\bigg|_\Gamma = \frac{D}{2 A_\mathrm{ex}} m_y,\\
        \frac{\partial m_x}{\partial x}\bigg|_\Gamma =  \frac{\partial m_y}{\partial y}\bigg|_\Gamma = -\frac{D}{2 A_\mathrm{ex}} m_z,\\
        \frac{\partial m_x}{\partial y}\bigg|_\Gamma = \frac{\partial m_y}{\partial x}\bigg|_\Gamma = 0.
    \end{gathered}
    \label{eq:boundary:conditions}
\end{equation}
\noindent
For example, when a cell $n$ lacks a neighboring cell $n \pm 1$ in the direction $\pm y$, the magnetization $\mathbf{m}_{n \pm 1}$ of this imaginary cell was defined as $\mathbf{m}_{n \pm 1} = \mathbf{m}_n \pm \displaystyle \frac{\partial \mathbf{m}}{\partial y} \bigg|_\Gamma l_y$, where $l_y$ is the cell size along the $y$ axis. Since the magnetization was assumed uniform in the thickness direction $z$ and the effective field $\mathbf{H}_\mathrm{eff}$ contained contributions resulting from PMA and DMI, there was no need in boundary conditions at the $\mathrm{W}|\mathrm{CoFeB}$ and $\mathrm{CoFeB}|\mathrm{MgO}$ interfaces.

The field-like and damping-like SOTs involved in Eq.~\eqref{eq:LLG} were calculated via the relations $\tau_{\mathrm{FL}(\mathrm{DL})}=\gamma \hbar (2 e M_s t_\mathrm{F})^{-1} \xi_{\mathrm{FL}(\mathrm{DL})} |J_x|$, where $\hbar$ is the reduced Planck constant, $e$ is the elementary positive charge, $J_x$ is the in-plane component of the 
electric current density $\mathbf{J}_\mathrm{W}$ in the $\mathrm{W}$ layer, and $\xi_\mathrm{FL}$ and $\xi_\mathrm{DL}$ are the coefficients depending on various parameters, such as the spin Hall angle of $\mathrm{W}$, thickness $t_\mathrm{W}$ of the $\mathrm{W}$ layer, spin-mixing conductance of the $\mathrm{W}|\mathrm{CoFeB}$ interface, and temperature~\cite{Paul2013}. In our numerical calculations, we used the coefficients $\xi_\mathrm{FL} = -0.0528$ and $\xi_\mathrm{DL} = -0.267$ experimentally measured for the $\mathrm{CoFeB}/\mathrm{W}$ bilayer with $t_\mathrm{W} = 5$~nm at room temperature~\cite{Changsoo2020}. However, the effect of nonzero field-like SOT was found to be negligible. We also assumed that the direct electric current flows entirely in the $\mathrm{W}$ layer, which is justified by a much higher conductivity of $\mathrm{W}$ ($1.8 \times 10^7$~S~m$^{-1}$~\cite{Hall1995}) in comparison with that of $\mathrm{CoFeB}$ ($4.4 \times 10^5$~S~m$^{-1}$~\cite{Fan2013}). It should be noted that the electric current injected into the $\mathrm{W}$ layer near its center $x = x_c$ (see Fig.~\ref{fig:structure}) has a position-dependent direction and density. Since this feature complicates the simulations without influencing the spin-wave propagation along the waveguide, we took into account only the in-plane component $J_x$ of the current density $\mathbf{J}_\mathrm{W}$ in the $\mathrm{W}$ layer and approximated its distribution as
\begin{equation}
    J_x = \begin{cases} -\displaystyle\frac{I_\mathrm{dc}}{2 w_\mathrm{F} t_\mathrm{W}} \; \text{at} \; x < x_c - d/2\\
                      0 \; \text{at} \; x_c - d/2 < x < x_c + d/2\\
                      \displaystyle\frac{I_\mathrm{dc}}{2 w_\mathrm{F} t_\mathrm{W}} \; \text{at} \; x > x_c + d/2
        \end{cases},
\end{equation}      
\noindent
where $w_\mathrm{F}$ is the waveguide width, while  $d$ is the injector width, which was assumed to be 400~nm. Below we will use the notation $J_\mathrm{W} = J_x(x>x_c+d/2).$

The numerical integration of Eq.~\eqref{eq:LLG} was performed using the Runge-Kutta projective algorithm with time step $\delta t = 10$~fs, which was found to be small enough to obtain a stable solution for the magnetization dynamics. We considered the $\mathrm{CoFeB}$ waveguide with the thickness $t_\mathrm{F} = 1.7$~nm, width $w_\mathrm{F} = 40$~nm, and length $L_\mathrm{F} = 6$~$\mu$m and divided it into a two-dimensional grid of computational cells with the dimensions $l_x =5$~nm, $l_y = 4$~nm, and $l_z = 1.7$~nm. An in-plane external magnetic field with the strength $H_y = 750$~Oe was introduced in the simulations, while additional Oersted fields created by the electric currents flowing in the heterostructure were neglected, because the calculations showed that they do not exceed 5\% of $H_y$ even at the current density of $5\times 10^{10}$~A~m$^{-2}$. Note that since $\mathbf{s} \times \mathbf{H} = 0$ no transverse relaxation of spin accumulation occurs in $\mathrm{W}$. The following values of the involved material parameters were employed in the numerical calculations: $M_s = 1.13 \times 10^6$~A~m$^{-1}$~\cite{Lee2011}, $A_\mathrm{ex} = 19$~pJ~m$^{-1}$~\cite{Sato2012}, $K_1 = 5 \times 10^3$~J~m$^{-3}$~\cite{Hall1960}, $K_s^0 = -1.3$~mJ~m$^{-2}$~\cite{Ikeda2010}, $k_s = 31$~fJ~V$^{-1}$~m$^{-1}$~\cite{Alzate2014}, $D = 0.42$~pJ~m$^{-1}$~\cite{Chaurasiya2016}, and $\alpha_0 = 0.01$~\cite{Ikeda2010}. Since the magnetization precession in the $\mathrm{CoFeB}$ layer leads to the spin pumping into the adjacent $\mathrm{W}$ film, the damping parameter $\alpha = \alpha_0 + \delta \alpha$ involved in Eq.~\eqref{eq:LLG} differs from the bulk Gilbert parameter $\alpha_0$ by the correction term $\delta \alpha\approx \displaystyle \frac{g_L \mu_B}{4 \pi M_s t_\mathrm{F}}\mathrm{Re}\big[g_r^{\uparrow \downarrow} \big]$~\cite{Tserkovnyak2002}, where $g_L$ is the Land\'{e} factor, $\mu_B$ is the Bohr magneton, and $g_r^{\uparrow \downarrow}$ is the complex reflection spin-mixing conductance per unit area of the $\mathrm{CoFeB}|\mathrm{W}$ interface. Using the experimentally determined value $\mathrm{Re}\big[g_r^{\uparrow \downarrow} \big] = 2.35$~nm$^{-2}$~\cite{Jhajhria2019}, we obtained $\alpha \approx 0.012$. 

\section{RESULTS OF MICROMAGNETIC MODELING}
\begin{figure}
    \center{\includegraphics[width=0.6\linewidth]{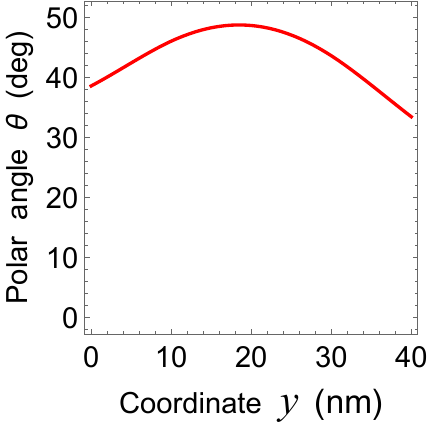}}
    \caption{\label{fig:profile} Equilibrium magnetization profile in the $\mathrm{CoFeB}$ waveguide. The plot shows the variation of the polar angle $\theta = \mathrm{arccos}(m_z)$ across the waveguide width at the applied magnetic field $H_y = 750$~Oe. The azimuthal angle $\phi = \mathrm{arctan}(m_y/m_x) \approx 90^\circ$ does not depend on the coordinate $y$. The magnetization profile remains almost the same along the waveguide length, changing significantly only near its ends. }
\end{figure}

First, we employed the micromagnetic simulations for the determination of the initial magnetization orientation in the $\mathrm{CoFeB}$ film. The study of the magnetization relaxation to an equilibrium direction showed that the magnetization is practically orthogonal to the $\mathrm{CoFeB}$ surfaces in the absence of external magnetic fields. This is due to the influence of PMA, which is stronger than that of the demagnetizing field $\mathbf{H}_\mathrm{dip}$ at the considered small $\mathrm{CoFeB}$ thickness $t_\mathrm{F} = 1.7$~nm. Under the external field $H_y = 750$~Oe, the magnetization rotates towards the film plane and becomes inhomogeneous across the waveguide (Fig.~\ref{fig:profile}), having an elevation angle of $49^\circ$ at the center. The revealed significant inhomogeneity of the magnetic state is caused by the demagnetizing field $\mathbf{H}_\mathrm{dip}$, which is rather strong owing to the relatively small width $w_\mathrm{F} = 40$~nm of the waveguide. The calculations also show that the asymmetry of the magnetization distribution across the waveguide (Fig.~\ref{fig:profile}) is due to the interfacial DMI.

\begin{figure}
    \center{\includegraphics[width=1.0\linewidth]{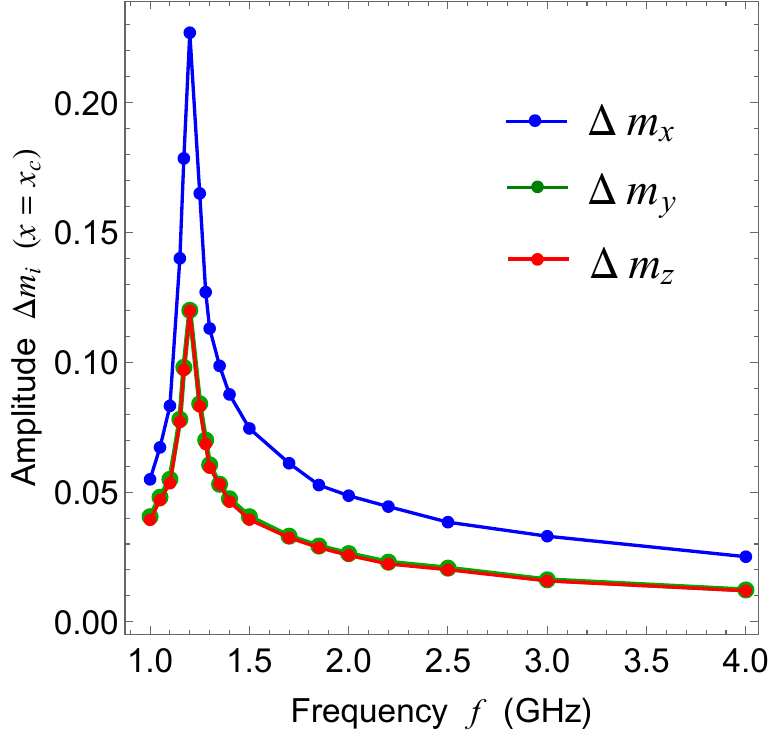}}
    \caption{\label{fig:resonance} Amplitude of the magnetization precession at the center of the $\mathrm{CoFeB}$ waveguide as a function of frequency $f$ of the microwave voltage $V_\mathrm{ac}$. The plots show the maximal changes $\Delta m_i$ of the magnetization direction cosines $m_i(x = x_c, y)$ averaged over the waveguide width. The voltage amplitude $V_\mathrm{max}$ equals 0.2~V. }
\end{figure}

Next, we studied the electrically induced magnetic dynamics in the waveguide in the absence of a direct electric current in the $\mathrm{W}$ layer. It was assumed that the top electrode with the dimensions $100 \times 40$~nm$^2$ deposited on the $\mathrm{MgO}$ layer with the thickness $t_\mathrm{MgO}=2$~nm is subjected to a microwave voltage $V_\mathrm{ac} = V_\mathrm{max} \sin{(2 \pi f t)}$. The voltage-induced modification $K_s = K_s^0+k_s V_\mathrm{ac}/t_\mathrm{MgO}$ of the PMA parameter associated with the $\mathrm{CoFeB}|\mathrm{MgO}$ interface beneath the electrode was taken into account for the corresponding computational cells. The calculations showed that the microwave voltage excites a steady-state magnetization precession $\delta m_i(t) = m_i(t) - m_i(t=0)$ ($i=x,y,z$) in the waveguide section under the top electrode. The quantities $\Delta m_i = \mathrm{max}[\langle \delta m_i(t) \rangle_y]$ characterizing the precession amplitude averaged over the waveguide width maximize when the excitation frequency $f$ equals $f_\mathrm{res} \approx 1.2$~GHz (see Fig.~\ref{fig:resonance}). At frequencies $f \geq f_\mathrm{res}$ the magnetization precession has an ellipticity $\epsilon \approx \Delta m_x / \sqrt{\Delta m_y^2+\Delta m_z^2}$, which is about 1.4 at the voltage amplitude $V_\mathrm{max} = 0.2$~V. It should be noted that the applied voltage $V_\mathrm{ac}(t)$ creates a microwave tunnel current flowing through the $\mathrm{MgO}$ barrier and the $\mathrm{CoFeB}/\mathrm{W}$ bilayer. Using the experimentally determined barrier conductance $G(t_\mathrm{MgO} = 2~\text{nm}) = 10^{7}$~S~m$^{-2}$~\cite{Yang2010}, we find that this current has a density $J_\mathrm{ac} \approx 2 \times 10^7$~A~m$^{-2}$ in the 5-nm-thick $\mathrm{W}$ layer at $V_\mathrm{max} = 0.2$~V. As confirmed by additional simulations, the SOT created by such an electric current does not significantly affect the magnetization dynamics in $\mathrm{CoFeB}$. 

\begin{figure}[h]
    \includegraphics[width=1\linewidth]{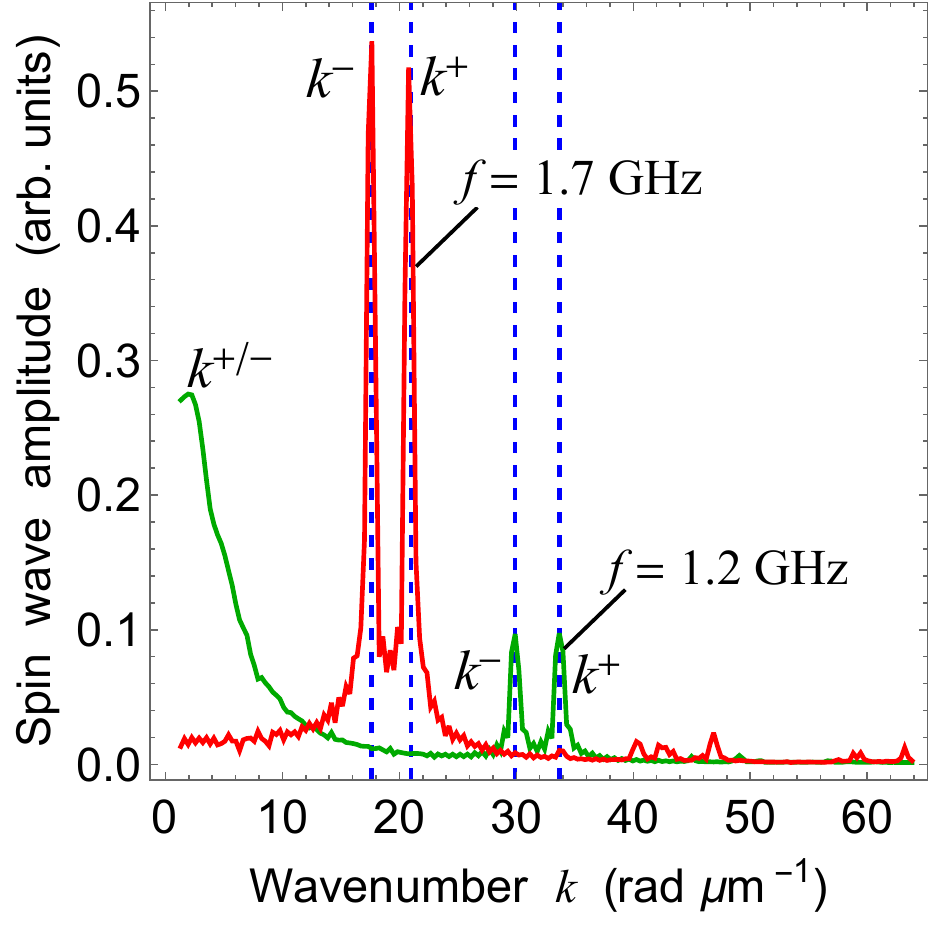}
\caption{\label{fig:spectr} Wavenumber spectra of spin waves generated by oscillating VCMA at excitation frequencies $f_\mathrm{res} = 1.2$~GHz and $f = 1.7$~GHz. While at $f = 1.7$~GHz the spin waves propagating in the waveguide have definite wavenumbers, a broad packet of modes with small wavenumbers also appears at the resonant excitation. Note that the presence of two separate peaks at $k_+$ and $k_-$ is due to nonreciprocal spin wave propagation caused by interfacial DMI. }
\end{figure}

\begin{figure}[h]
    \center{
    \includegraphics[width=1.0\linewidth]{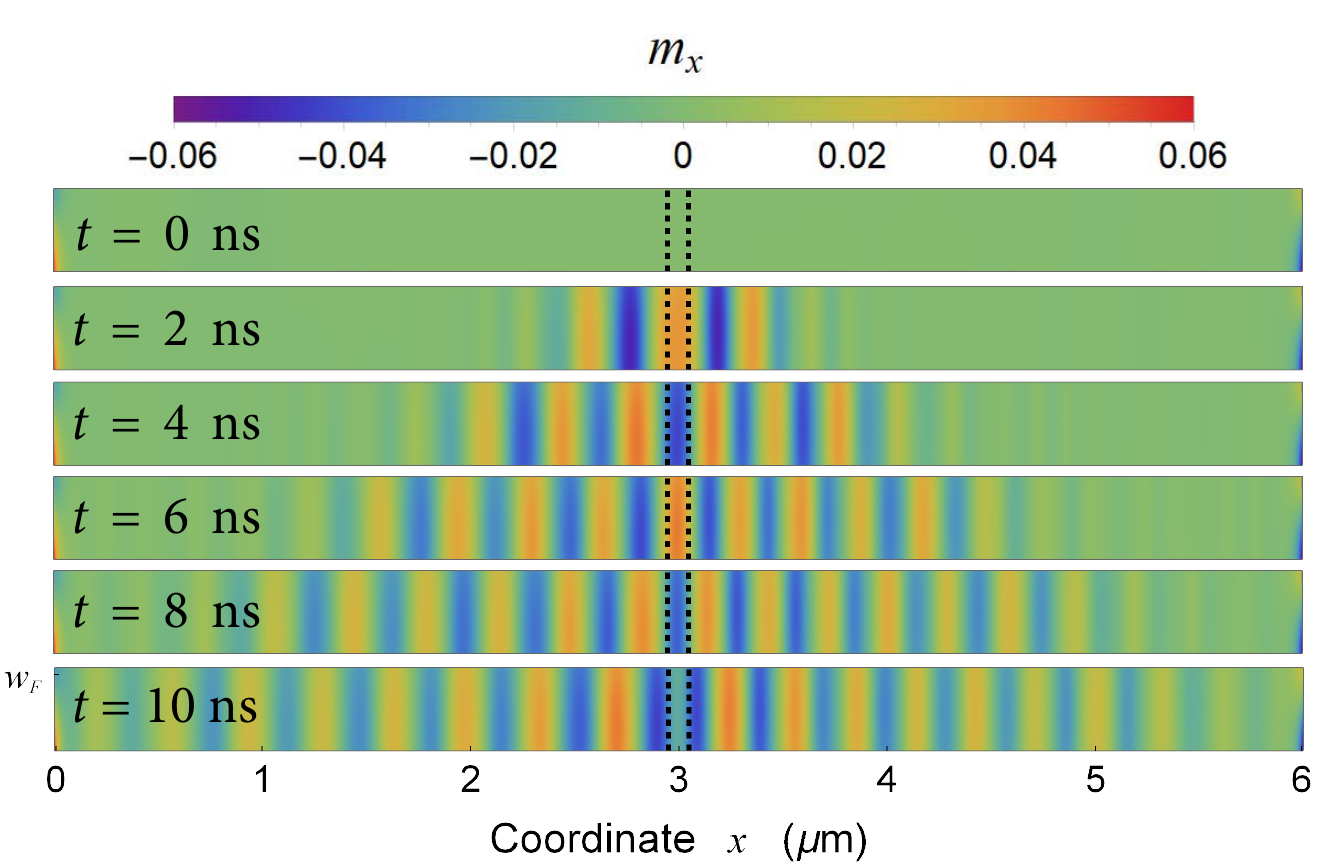}
    }
    \caption{\label{fig:sketches} Time evolution of spin waves generated in the center of the $\mathrm{CoFeB}$ waveguide. Color diagrams show distributions of the magnetization direction cosine $m_x(x, y)$ in the waveguide at different moments of time. Dashed lines mark the boundaries of the excitation area, where the ac voltage with the frequency $f = 1.7$~GHz and amplitude $V_\mathrm{max} = 0.2$~V is applied to the $\mathrm{MgO}$ nanolayer. }
\end{figure}

At the excitation frequencies below the resonance frequency $f_\mathrm{res}$, the electrically induced magnetization precession appears to be confined within the waveguide section under the top electrode. In contrast, the emission of spin waves from the excitation area was revealed at the frequencies $f \geq f_\mathrm{res}$. These waves travel in the opposite directions within two halves of the $\mathrm{CoFeB}$ nanolayer. In the case of resonant excitation ($f \approx f_\mathrm{res}$), packets of spin-wave modes with various wave vectors propagate in the waveguide (see Fig.~\ref{fig:spectr}). However, well above $f_\mathrm{res}$ the magnetization dynamics takes the form of a spin wave with a definite wavenumber $k_x(x>x_c) = k_+$ or $k_x(x<x_c) = k_-$ (Fig.~\ref{fig:spectr}). Therefore, further modeling was carried out at the excitation frequency $f = 1.7$~GHz, at which the simulations yield $k_+ = 21.08$~rad~$\mu$m$^{-1}$ and $k_- = 17.42$~rad~$\mu$m$^{-1}$. Figure~\ref{fig:sketches} illustrates time evolutions of the two spin waves generated by the microwave voltage with such a frequency. Interestingly, despite significant difference between the wavenumbers $k_+$ and $k_-$, the amplitudes of the two spin waves decrease very similarly with the distance $|x - x_c|$ from the waveguide center. The decay of the spin-wave amplitude, which is caused by the Gilbert damping, follows the exponential law $\Delta m_x(x) = \Delta m_x(x_c) \exp{[-|x-x_c| / \lambda_\pm]}$ with a high accuracy. The decay lengths $\lambda_+$ and $\lambda_-$ of spin waves with the wavenumbers $k_+$ and $k_-$ were found to be $\lambda_+ \approx \lambda_- \approx 2.5$~$\mu$m. It should be noted that the revealed difference between $k_+$ and $k_-$ is due to the interfacial DMI. Such a difference was earlier predicted theoretically~\cite{Moon2013} and observed experimentally in the $\mathrm{W}/\mathrm{CoFeB}/\mathrm{SiO}_2$ heterostructure~\cite{Chaurasiya2016}.

\begin{figure}
    \center{\includegraphics[width=0.9\linewidth]{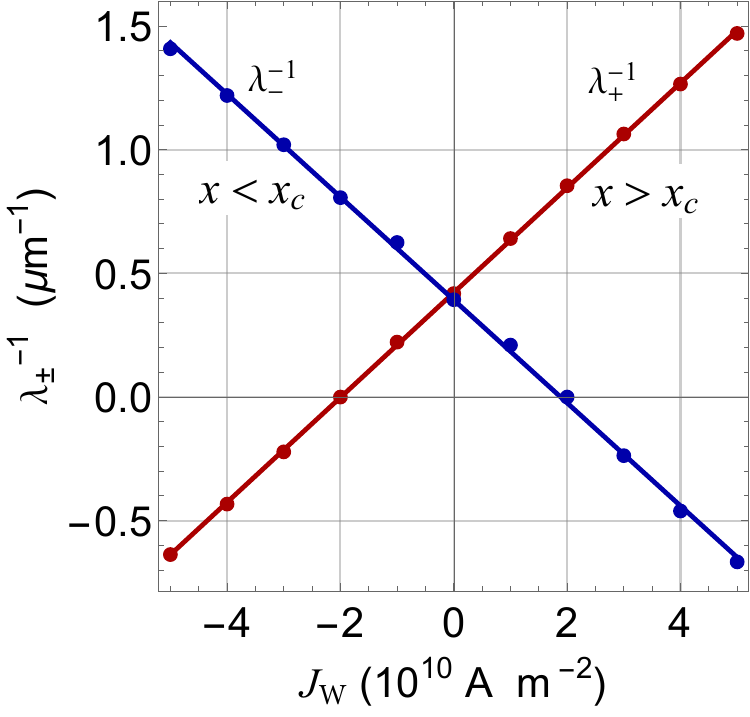}}
    \caption{\label{fig:decay} Variations of inverse decay lengths $1 / \lambda_+$ and $1 / \lambda_-$ of spin waves propagating in the $\mathrm{CoFeB}$ waveguide with the density $J_\mathrm{W}$ of direct electric current flowing in the $\mathrm{W}$ film.  Points show the inverse decay lengths extracted from the simulation data, and lines represent linear fits of the results obtained for the waves with the wave vectors $k_+$ and $k_-$ travelling at $x > x_c$ and $x < x_c$, respectively. }
\end{figure}

The most important results were obtained when studying the influence of the direct electric current flowing in the $\mathrm{W}$ layer on the propagation of spin waves in the $\mathrm{CoFeB}$ waveguide. For each predetermined density $J_\mathrm{W}$ of this current, we modeled the excitation of spin waves by the microwave voltage $V_\mathrm{ac}(t)$ locally modulating VCMA in the presence of SOT created by the spin current injected into the CoFeB nanolayer. In the simulations, the VCMA oscillating with the frequency $f = 1.7$~GHz $> f_\mathrm{res}$ and time-independent SOT were simultaneously applied at the moment $t = 0$ to the CoFeB nanolayer with the equilibrium magnetization distribution forming at $V_\mathrm{ac} = 0$ and $J_\mathrm{W} = 0$. The analysis of the electrically induced magnetization dynamics was limited by the time period $t \simeq 5$~ns, during which both spin waves reach the ends of the $6$-$\mu$m-long waveguide. It was found that SOT does not significantly change the wave numbers $k_+$ and $k_-$ of the generated spin waves. At small current densities $J_\mathrm{W}$, the amplitudes of these waves still decrease exponentially with the increasing distance $|x - x_c|$ from the waveguide center. However, their decay lengths $\lambda_+$ and $\lambda_-$ change under the action of SOT in the opposite way owing to different directions of the electric currents flowing in the underlying halves of the $\mathrm{W}$ layer (Fig.~\ref{fig:structure}), which affect the vector $\mathbf{s}$ in Eq.~\eqref{eq:LLG}. Namely, the spin wave decays faster when the quantity $\tau_\mathrm{DL} \int_{t}^{t + 1/f} dt \big[(\mathbf{m} \cdot \mathbf{H}_\mathrm{eff})(\mathbf{m} \cdot \mathbf{s}) - \mathbf{s} \cdot \mathbf{H}_\mathrm{eff} \big]$ is positive and slower when it is negative. The simulation data show that the inverse decay lengths $1 / \lambda_+$ and $1 / \lambda_-$ vary linearly with the current density $J_\mathrm{W}$ (see Fig.~\ref{fig:decay}).

Remarkably, $1 / \lambda_+$ and $1 / \lambda_-$ go to zero at $J_\mathrm{W}^+ \approx -2 \times 10^{10}$~A~m$^{-2}$ and $J_\mathrm{W}^- \approx 2 \times 10^{10}$~A~m$^{-2}$, respectively. Hence, the spin wave with the wave vector $\mathbf{k}^+$ ($\mathbf{k}^-$) travels with a constant amplitude at the critical current density $J_\mathrm{W}^+$ ($J_\mathrm{W}^-$) because the Gilbert damping is fully compensated by SOT. At overcritical current densities $J_\mathrm{W} < J_\mathrm{W}^+$ or $J_\mathrm{W} > J_\mathrm{W}^-$, the amplitude of such a spin wave exponentially increases with the distance from the excitation area, which manifests itself in negative values of the decay lengths seen in Fig.~\ref{fig:decay}. 

\begin{figure}
    \center{\includegraphics[width=0.9\linewidth]{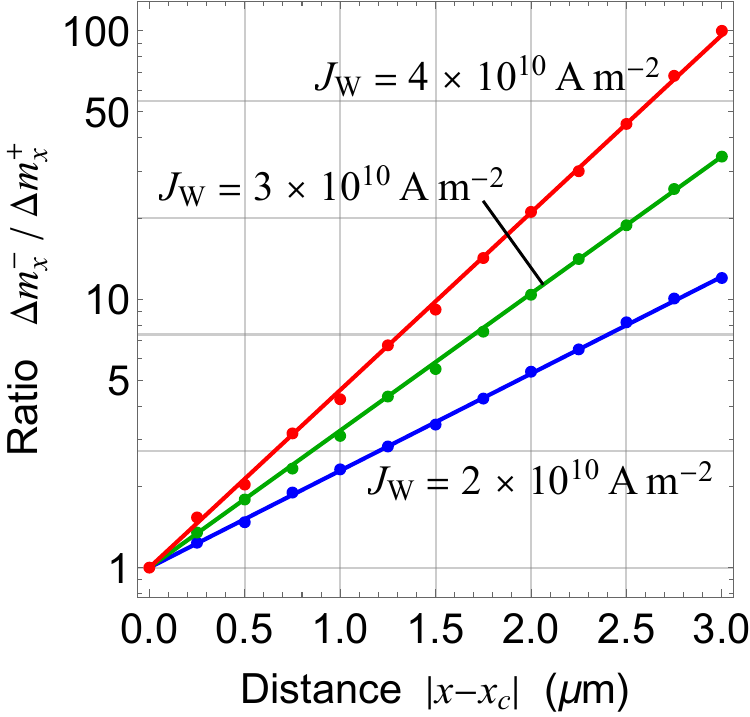}}
    \caption{\label{fig:AMPratio} Ratio $\Delta m_x^-/\Delta m_x^+$ of the precession amplitudes in the spin waves with the wavenumbers $k_-$ and $k_+$ plotted as a function of distance $|x - x_c|$ from the waveguide center. Points represent the simulation data, which are fitted by the exponential law (lines). }
\end{figure}

Since SOT simultaneously reduces the positive decay length of another spin wave that travels in the opposite direction, by passing sufficient direct current through the $\mathrm{W}$ layer it becomes possible to realize a long-distance spin-wave propagation in one half of the $\mathrm{CoFeB}$ waveguide only. Moreover, by reversing the polarity of the dc voltage applied to the $\mathrm{W}$ layer one can change the propagation region and switch the travel direction of the spin wave in the $\mathrm{CoFeB}$ waveguide. To quantify the described effect, we determined the ratio $\Delta m_x^- / \Delta m_x^+$ of the precession amplitudes in the spin waves with the wavenumbers $k_-$ and $k_+$ at the same distance $|x - x_c|$ from the waveguide center. Variations of this ratio with the distance $|x - x_c|$ at different densities $J_\mathrm{W} \geq J_\mathrm{W}^-$ of the electric current are presented in Fig.~\ref{fig:AMPratio}. 
It is seen that the ratio $\Delta m_x^- / \Delta m_x^+$ exponentially increases with the distance from the source of spin waves, reaching 100 at $|x - x_c| = 3$~$\mu$m when $J_\mathrm{W} = 4 \times 10^{10}$~A~m$^{-2}$. Since at $f = 1.7$~GHz the propagation of monochromatic spin waves with definite wavenumbers takes place in the waveguide (see Fig.~\ref{fig:spectr}), the ratio of the amounts of power transmitted in the opposite directions at the distance  $|x - x_c|$ simply equals the square of the corresponding number given in Fig.~\ref{fig:AMPratio}.  

To clarify the optimal conditions for the SOT-induced amplification of spin waves in the $\mathrm{CoFeB}$ waveguide, we studied the dependence of the critical current densities $J_\mathrm{W}^+$ and $J_\mathrm{W}^-$ on the strength $H_y$ of the external in-plane magnetic field. 
\begin{figure}[!h]
    \centering
    \includegraphics[width=0.8\linewidth]{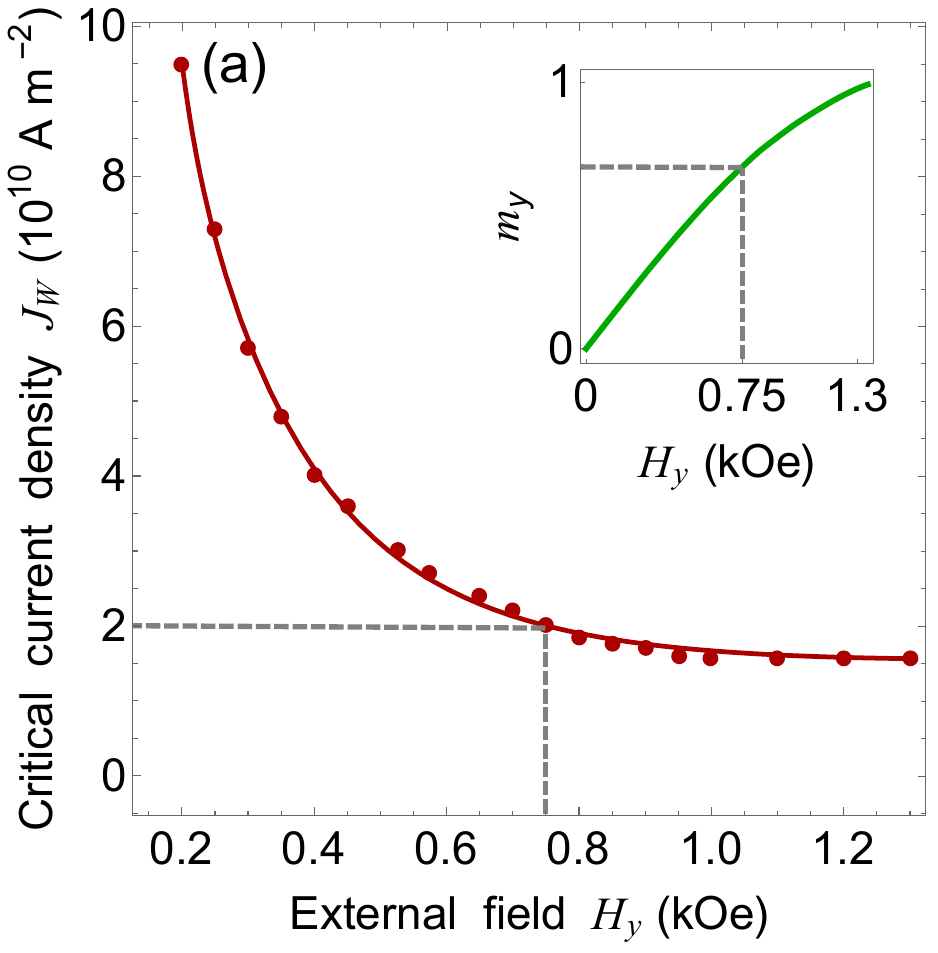}
    \includegraphics[width=0.8\linewidth]{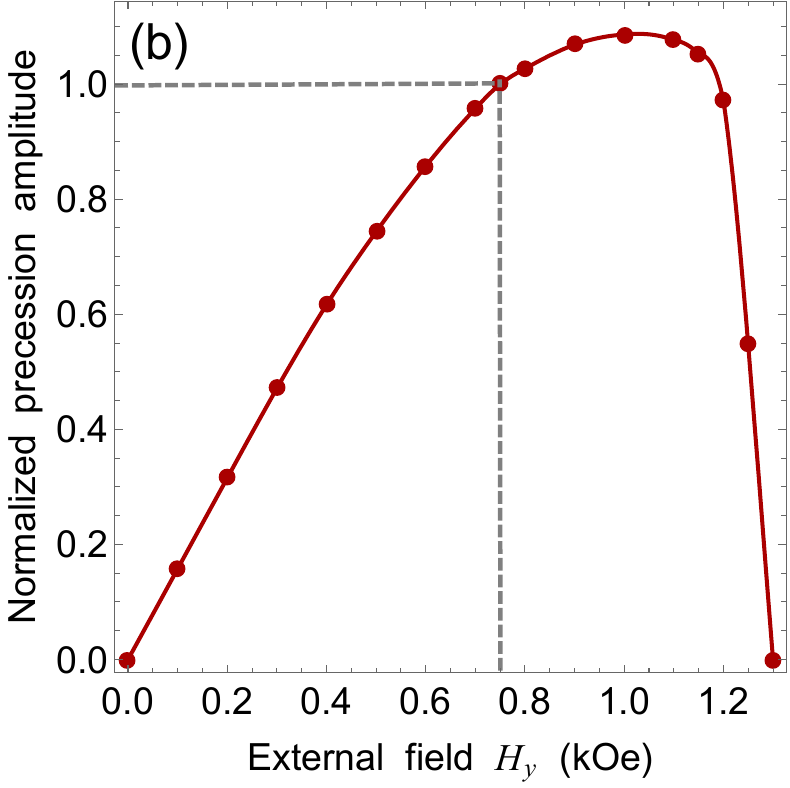}
   \caption{\label{fig:efficiency} Influence of external in-plane magnetic field on the critical current density in W layer (a) and the resonant magnetization precession in CoFeB waveguide (b). Data points indicate the absolute value of the critical densities $J_\mathrm{W}^+ \approx -J_\mathrm{W}^-$ and the precession amplitude at $f_\mathrm{res}$($H_y$) normalized by its value at $H_y = 750$~Oe. The inset in panel (a) shows the magnetic-field dependence of the in-plane component $m_y$ of the equilibrium magnetization averaged over the waveguide width. Dashed lines mark the results obtained in the simulations performed at $H_y = 750$~Oe. }
\end{figure}
The simulations showed that the magnitude of critical densities monotonically decreases with increasing field strength [see Fig.~\ref{fig:efficiency} (a)]. However, this decrease is accompanied by a gradual rotation of the equilibrium magnetization direction in the $\mathrm{CoFeB}$ layer towards the in-plane orientation [see inset in Fig.~\ref{fig:efficiency} (a)]. Owing to such a rotation, the amplitude of magnetization precession in the waveguide varies nonmonotonically with increasing $H_y$. As demonstrated by Fig.~\ref{fig:efficiency} (b), the precession amplitude becomes maximal at $H_y \approx 1$~kOe  and decreases drastically when the field strength exceeds 1.2~kOe. Therefore, the applied magnetic field should not be higher than 1.2 kOe so that the lowest acceptable magnitude of the critical current densities amounts to about $ 1.6 \times 10^{10}$~A~m$^{-2}$. This result shows that the chosen field strength $H_y = 750$~Oe provides almost minimal critical current density in the W layer in addition to almost maximal precession amplitude in the $\mathrm{CoFeB}$ waveguide.

Discussing our theoretical results in the light of available experimental data, we note that a strong effect of SOT on the propagation of spin waves was observed in the Yttrium Iron Garnet (YIG) waveguide~\cite{Evelt2016}. It was found that the spin-wave decay length could be increased by nearly a factor of 10 by passing an electric current through the $\mathrm{Pt}$ layer adjacent to the YIG film. Up to some threshold current, the inverse decay length decreases linearly with the current magnitude, but then begins to increase instead of tending to zero~\cite{Evelt2016}. The absence of the expected compensation of the magnetic damping by SOT was attributed to the enhancement of magnetic fluctuations by SOT, which becomes important at large currents~\cite{Demidov2020}. In the YIG$/\mathrm{Pt}$ system, however, the estimated current density providing the damping compensation is about 16 times larger than the critical density $J_\mathrm{W} \approx 2 \times 10^{10}$~A~m$^{-2}$ that we predict for the current flowing in the $\mathrm{W}$ layer of the $\mathrm{W}/\mathrm{CoFeB}/\mathrm{MgO}$ heterostructure. Furthermore, the maximal amplitude $\Delta m_x$ of the magnetization precession in our simulations does not exceed 0.15, and the precession ellipticity is about 1.4 only. These values explain the SOT-induced amplification of spin waves, which was revealed in our simulations during a short time period $t = 6$~ns in the absence of thermal fluctuations. Indeed, recent study of the magnetization dynamics in ferromagnetic disks demonstrated that the nonlinear damping could be suppressed by minimizing the precession ellipticity even at amplitudes exceeding 0.15~\cite{Divinskiy2019}. 

However, a longer modeling demonstrates the progressive development of magnetization auto-oscillations in the CoFeB layer at the current densities $J_\mathrm{W} \leq J_\mathrm{W}^+$ and $J_\mathrm{W} \geq J_\mathrm{W}^-$. The amplitude of these oscillations strongly increases at the time $t \simeq 20 \div 30$~ns after the application of SOT, which leads to the transfer of the spin-wave power to incoherent auto-oscillations with higher frequencies due to the magnon-magnon scattering. As a result, the spin-wave amplitude now decreases with the distance from the excitation area even at the highest current density $J_\mathrm{W} = 5 \times 10^{10}$~A~m$^{-2}$ used in our simulations. Thus, the compensation of the pronounced nonlinear damping by SOT appears to be impossible~\cite{Demidov2020}, but the SOT-induced amplification of spin waves during several nanoseconds should be feasible in the W/CoFeB/MgO nanostructure. This prediction is corroborated by the recent experimental demonstration of the spin-wave amplification by short SOT pulses in the YIG/Pt waveguide~\cite{Demidov2023}.

To check whether the temporary amplification of spin waves in the W/CoFeB/MgO nanostructure is also possible in the presence of thermal fluctuations, we carried out additional simulations, in which the effective field $\mathbf{H}_\mathrm{eff}$ involved in Eq.~\eqref{eq:LLG} was appended by the contribution of a thermal random field $\mathbf{H}_\mathrm{th}$~\cite{Ito2006}. The simulations were performed using MuMax3 software~\cite{mumax3} for the 30-$\mu$m-long waveguide kept at the temperature $T = 300$~K. To increase the stability of the out-of-plane magnetization state against deflections caused by thermal fluctuations, the CoFeB width $w_\mathrm{F}$ was increased from 40 to 100~nm. 
\begin{figure*}[!httb]
\centering
\includegraphics[width = 0.3\linewidth]{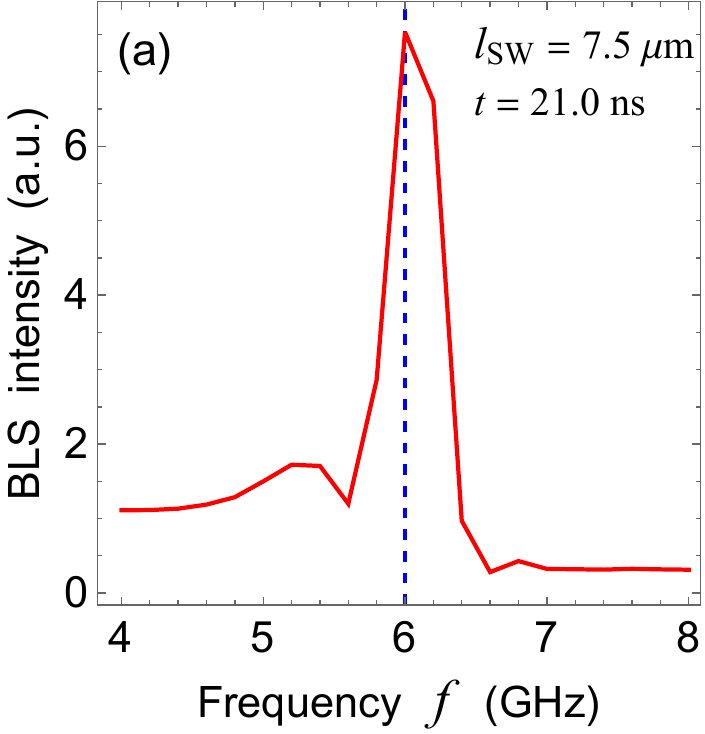}
\hfill
\includegraphics[width = 0.3\linewidth]{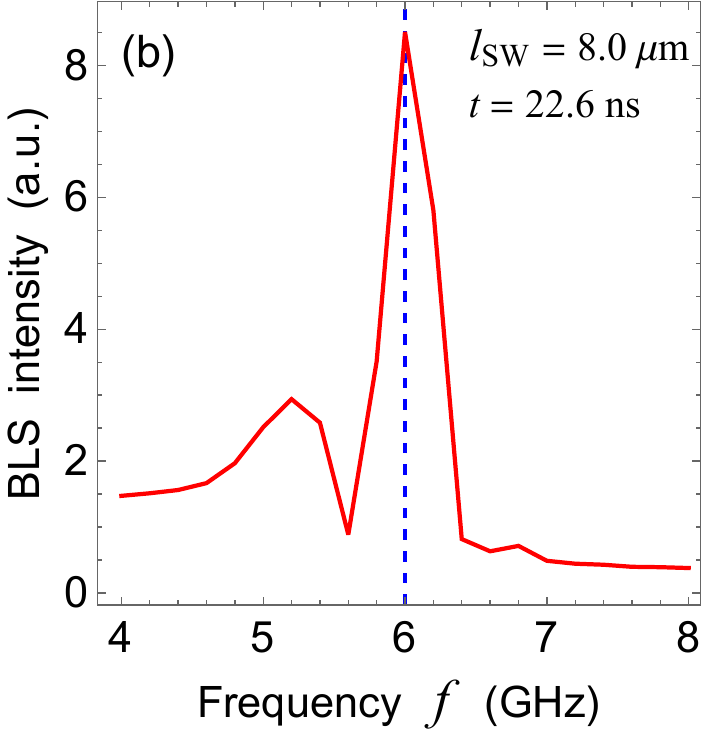}
\hfill
\includegraphics[width = 0.3\linewidth]{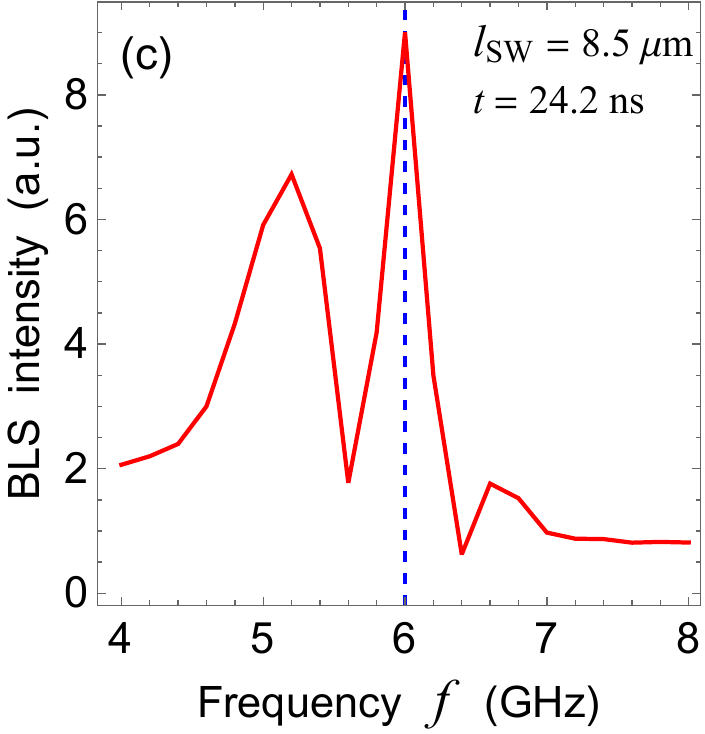}
\caption{\label{fig:BLS-f} Frequency spectra of the BLS intensity calculated for the waveguide sections with the in-plane sizes $\delta x = 100$~nm and $\delta y = w_\mathrm{F} = 100$~nm situated at the positions $x_\mathrm{SW}(t)$ of the spin-wave front in the CoFeB waveguide. The spin wave propagates in the left half of the waveguide, and the distance $l_\mathrm{SW}(t)$ between its front and the waveguide center equals $l_\mathrm{SW}(t = 21 \;\text{ns}) = 7.5$~$\mu$m (a), $l_\mathrm{SW}(t = 22.6 \;\text{ns}) = 8$~$\mu$m (b), and $l_\mathrm{SW}(t = 24.2 \;\text{ns}) = 8.5$~$\mu$m (c). The density of the electric current in the W layer is set to $J_\mathrm{W} = 20 \times 10^{10}$~A~m$^{-2}$. }
\end{figure*}
An applied magnetic field with the components $H_y = 540$~Oe and $H_z = 1700$~Oe was introduced to ensure efficient nonparametric generation of spin waves by the modulation of VCMA at the frequencies above the resonance frequency, which amounts to 5.1~GHz at such a field. In addition, the periodic boundary conditions $\mathbf{m}(x = L_\mathrm{F}) = \mathbf{m}(x = 0)$ were employed to get rid of spin-wave reflections at the waveguide ends. 

The simulations executed for a representative excitation frequency $f = 6$~GHz and various densities $J_\mathrm{W} \sim 10^{11}$~A~m$^{-2}$ of the electric current in the W layer demonstrated the generation of both monochromatic spin waves and incoherent magnetization oscillations in the CoFeB layer. Since the CoFeB thickness is very small, the influence of thermal fluctuations on the magnetic dynamics appears to be strong at $T = 300$~K. Therefore, the quantification of the spin-wave amplitude requires some averaging procedure. Since the propagation of spin waves in ferromagnets can be visualized with the aid of the micro-focus Brillouin light scattering (BLS) spectroscopy~\cite{Demidov2015}, we calculated the BLS intensity corresponding to the magnetization dynamics predicted by our simulations. Using the computational technique described in Ref.~\cite{Wojewoda2023}, we obtained frequency spectra of the BLS intensities at the fronts $x_\mathrm{SW}(t)$ of two propagating spin waves at different times $t$ after their generation in the waveguide center. The representative spectra presented in Fig.~\ref{fig:BLS-f} demonstrate the existence of two pronounced peaks, one of which is centered around the excitation frequency $f = 6$~GHz, whereas the other is located at a frequency $f = 5.2$~GHz close to the resonance frequency $f\mathrm{res} = 5.1$~GHz of the CoFeB layer. Since the background BLS intensity independent of frequency and spatial location was subtracted from the spectra shown in Fig.~\ref{fig:BLS-f}, the height of the first peak quantifies the intensity of the spin wave at the position $x_\mathrm{SW}(t)$ of its front. In contrast, the second peak characterizes the magnetization auto-oscillations developed during the time period $t$ of the spin-wave-propagation in the waveguide. The comparison of the three spectra presented in Fig.~\ref{fig:BLS-f} shows that the intensity of such oscillations rapidly grows with time and approaches the spin-wave intensity after about 24~ns.

\begin{figure}
    \center{\includegraphics[width=\linewidth]{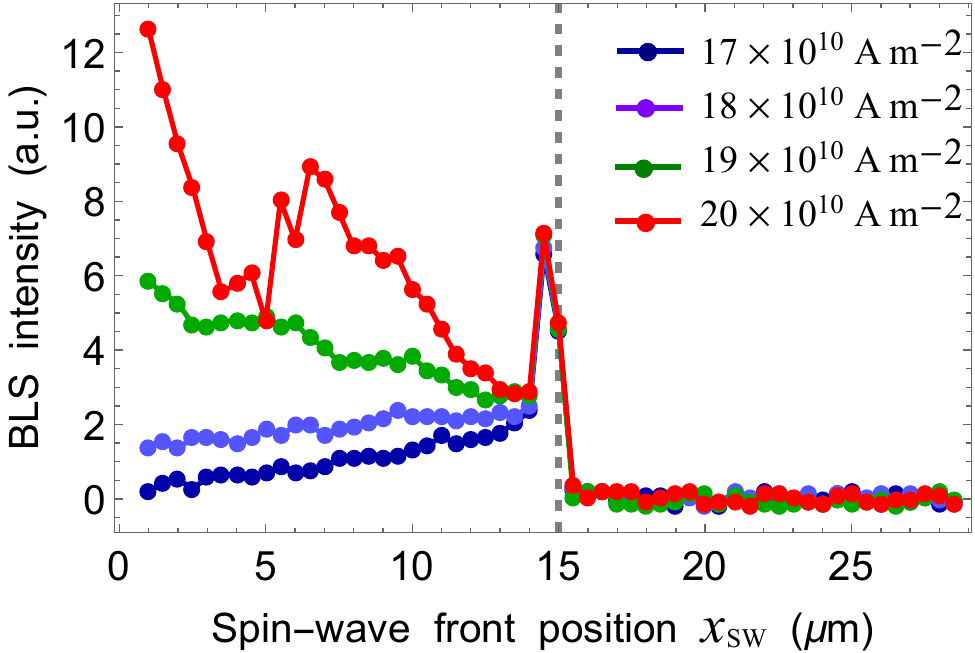}}
    \caption{\label{fig:BLS-x} Variations of the BLS intensity calculated at the frequency $f = 6$~GHz on the position $x_\mathrm{SW}$ of the spin-wave front in the waveguide. Values of the density $J_\mathrm{W}$ of the electric current in the W layer are indicated in the figure. The dashed line shows the position of the excitation area in the center of the waveguide. }
\end{figure}

Representative dependences of the BLS intensity at the frequency $f = 6$~GHz on the position $x_\mathrm{SW}(t)$ of the spin-wave front in the waveguide are shown in Fig.~\ref{fig:BLS-x}. Remarkably, they demonstrate that, when the current density $J_\mathrm{W}$ in the heavy-metal layer is high enough ($J_\mathrm{W} = 20 \times 10^{10}$~A~m$^{-2}$), the intensity of the spin wave propagating in the left half of the waveguide significantly increases with the distance $|x_\mathrm{SW}(t) - L_\mathrm{F} / 2|$ from the excitation area. Thus, the temporal amplification of spin waves by SOT should be feasible in the W/CoFeB/MgO nanostructure even at room temperature.

\section{Conclusion}
In this work, we theoretically studied the electrical excitation and control of spin waves in a ferromagnetic waveguide. The study was carried out for the $\mathrm{W}/\mathrm{CoFeB}/\mathrm{MgO}$ nanostructure having PMA and DMI associated with the $\mathrm{CoFeB}|\mathrm{MgO}$ and $\mathrm{W}|\mathrm{CoFeB}$ interfaces, respectively. Using micromagnetic simulations based on the numerical solution of the modified Landau-Lifshitz-Gilbert equation, we showed that the modulation of PMA by a microwave voltage locally applied to the $\mathrm{MgO}$ layer renders it possible to generate two spin waves propagating in the opposite directions from the center of the $\mathrm{CoFeB}$ waveguide. Although owing to DMI these waves have different wavenumbers (21.08~rad~$\mu$m$^{-1}$ vs. 17.42~rad~$\mu$m$^{-1}$ at $f = 1.7$~GHz), they retain similar decay lengths of about 2.5~$\mu$m. It should be noted that, in contrast to the previous theoretical works \cite{Verba2014, Verba2017}, the VCMA-driven excitation of spin waves is not parametric in our simulations. Due to the inclined magnetization direction in the ferromagnetic layer (Fig.~\ref{fig:structure}), the spin-wave generation takes place even at small amplitudes of the microwave voltage unlike the parametric excitation, which requires the voltage amplitude exceeding some threshold value \cite{Verba2014, Verba2017}. Furthermore, the simulations demonstrated that the propagation lengths of spin waves in the $\mathrm{CoFeB}$ layer can be changed drastically by passing a direct electric current through the adjacent $\mathrm{W}$ film. Depending on the direction of the electric current and that of the effective field $\mathbf{H}_\mathrm{eff}$, the propagation length either increases or decreases due to the SOT acting on the magnetization. Importantly, complete compensation of the magnetic damping by the SOT-induced antidamping temporarily occurs at the critical current density, which amounts to about $2 \times 10^{10}$~A~m$^{-2}$ at $T = 0$~K and $19 \times 10^{10}$~A~m$^{-2}$ at $T = 300$~K. This remarkable feature of the $\mathrm{W}/\mathrm{CoFeB}/\mathrm{MgO}$ heterostructure opens the possibility of efficient electrical control of the spin-wave propagation in the $\mathrm{CoFeB}$ layer. Hence, this nanostructure represents a promising waveguide, in which the amplification and long-range propagation of spin waves could be achieved in practice.

If the electric current is locally injected into the heavy-metal layer near the waveguide center, the charge flow has opposite directions in the two halves of this layer (Fig.~\ref{fig:structure}). Therefore, the spin waves travelling in the adjacent halves of the waveguide experience the action of SOTs having opposite directions. Because of such SOT inhomogeneity, it becomes possible to strongly increase the propagation length of one of these waves, while simultaneously creating a fast decay of the other wave. The simulations show that at $T = 0$~K the ratio of the amplitudes of magnetization precession at the two ends of the 6-$\mu$m-long $\mathrm{CoFeB}$ waveguide reaches 100 at the current density $J_\mathrm{W} = 4 \times 10^{10}$~A~m$^{-2}$. Hence, significant spin-wave signal can be sent to one of the waveguide ends only. Moreover, the signal transmission can be switched to the other end by changing the polarity of the dc voltage applied to the heavy-metal layer. Thus, the studied nanostructure $\mathrm{W}/\mathrm{CoFeB}/\mathrm{MgO}$ represents an electrically controlled magnonic device that converts the electrical input signal into a spin signal, which can be transmitted to one of two outputs. Since such a device does not employ oscillating magnetic fields, it may have relatively low power consumption, which should facilitate its applications in magnonics. \\

\section{ACKNOWLEDGMENT}
We thank Dr. Vladislav E. Demidov for valuable comments on our micromagnetic simulations of the spin-wave amplification in the presence of spin-orbit torque.

\bibliography{ref}
\end{document}